\documentclass{iau}

\makeatletter

\def\ArtDir#1{}
\def\logo@dir{}
\renewcommand{\orcid}[1][]{}
\makeatother

\usepackage{graphicx}

\usepackage{amsmath}
\usepackage{graphicx}
\usepackage{multirow}
\usepackage{csquotes}
\usepackage{natbib}

\begin{document}

\lefttitle{Dasgupta}
\righttitle{Technosignature and UHE Neutrino Synergies}

\jnlPage{1}{7}
\jnlDoiYr{2026}
\doival{10.1017/xxxxx}

\aopheadtitle{Proceedings IAU Symposium}
\editors{J. Haqq-Misra \& R. Kopparapu, eds.}

\title{Methodological Synergies between Technosignature and UHE Neutrino Searches}

\author{Paramita Dasgupta}
\affiliation{Department of Physics, Center for Cosmology and AstroParticle Physics, The Ohio State University, Columbus, OH 43210, USA}

\begin{abstract}
Radio technosignature searches and radio-based ultra-high energy (UHE; $E_{\nu} \geq 10^{17}$~eV) neutrino experiments address different scientific questions, but they share
a closely related data analysis problem. Both fields aim to identify rare signals of
poorly constrained or entirely unknown morphology within large datasets
dominated by thermal noise, anthropogenic radio-frequency interference (RFI),
instrumental artifacts, and poorly characterized backgrounds. UHE neutrino
radio experiments---including the in-ice Askaryan Radio Array (ARA) at the South Pole, the Radio Neutrino Observatory in Greenland (RNO-G), and the balloon-borne Antarctic Impulsive Transient Antenna (ANITA) and its successor the Payload for Ultrahigh Energy Observations (PUEO) over Antarctica---have developed advanced methodologies for continuous-wave (CW) mitigation, background characterization, and multi-stage candidate review through years of operating in remote yet radio-contaminated environments. 

This invited contribution makes that connection concrete through three
contributions. First, we demonstrate that catalog-based time-domain sine
subtraction---the CW mitigation technique used in ANITA and ARA
analyses---can be adapted for technosignature pipelines by restricting
subtraction to independently documented persistent contaminants, thereby
simultaneously improving broadband transient visibility and preserving
uncataloged narrowband candidates. Second, we identify a structural
equivalence, to the author's knowledge not previously stated, between the
spatiotemporal clustering used in UHE-neutrino experiments to
reject anthropogenic events and the direction- and cadence-based RFI
rejection used in radio SETI, and propose extending this to a joint feature space incorporating direction, time, frequency, bandwidth, duration, and
polarization. Third, we argue that background-only anomaly ranking is the
natural second stage of this workflow, providing morphology-agnostic candidate triage sensitive to signal classes no supervised template would
recover. Together these ideas motivate a \emph{preserve-then-rank} workflow
for commensal rare-event discovery that requires no modification to
either community's primary science pipeline and opens a concrete, near-term path toward cross-community collaboration in the shared challenge of rare-signal discovery.
\end{abstract}


\maketitle

\section{Introduction}
\label{sec:intro}

The search for technosignatures began with the suggestion by \citet{cocconi1959} that narrowband radio emission near the hydrogen line could serve as a natural interstellar communication channel. Project Ozma soon followed as the first systematic radio SETI experiment \citep{drake1961}. Since then, the exoplanet revolution has expanded the plausible target set \citep{dressing2015}, while the broader technosignature framing has extended the field beyond intentional communication toward a wider class of remotely detectable technological phenomena \citep{wright2018nomenclature,wright2022case}. Radio has remained central because it supports a wide range of plausible signal classes---narrowband, drifting, pulsed, and broadband---and modern instruments can survey wide frequency ranges efficiently \citep{tarter2001,worden2017,enriquez2017,gajjar2021}. The primary challenge today is no longer sensitivity alone. It is the need to identify rare and possibly unconventional candidates within large datasets dominated by radio-frequency interference (RFI), instrumental artifacts, and background structure.

A similar challenge motivates the design of radio ultra-high energy (UHE) neutrino experiments ($E_{\nu} \geq 10^{17}$~eV). More than a century after Victor Hess discovered cosmic rays, the astrophysical engines capable of accelerating particles to such extreme energies remain a mystery. Observatories like the Fly's Eye---famous for catching the $3 \times 10^{20}$~eV ``Oh-My-God'' particle \citep{bird1993}---and the Pierre Auger Observatory \citep{aab2020} have recorded cosmic rays with energies tens of millions of times higher than any human-made accelerator can produce. However, because these UHECRs, which consist of protons and heavier atomic nuclei, are electrically charged, their trajectories are scrambled by interstellar and intergalactic magnetic fields. This deflection completely obscures their distant origins. Neutrinos, by virtue of being electrically neutral and weakly interacting, travel in straight lines across the cosmos and are completely unaffected by these magnetic fields. Hence, they serve as unique ``cosmogenic'' messengers from the distant universe, providing us with a tool to probe deep into opaque sources to study cosmic acceleration mechanisms and the cosmological evolution of the universe.

A guaranteed flux of UHE neutrinos is anticipated when UHECRs ($E_{\rm CR} \gtrsim 10^{18}$~eV) interact with cosmic microwave background (CMB) and extragalactic background light (EBL) photons via the $\Delta$-resonance. This phenomenon, known as the Greisen-Zatsepin-Kuzmin (GZK) suppression \citep{greisen1966,zatsepin1966}, limits the horizon of UHECR propagation while producing cosmogenic neutrinos that can constrain the mass composition and source evolution of UHECRs across cosmological distances. However, detecting these UHE neutrinos is exceptionally challenging, requiring immensely large detector volumes for detecting a UHE neutrino event. At these extreme energies, the predicted flux and small interaction cross-sections yield roughly $\mathcal{O}(10^{-2})$ neutrino interactions per cubic kilometer of ice per year, turning UHE neutrino astronomy into a rare-event search that requires teraton-scale detectors. 

When a UHE neutrino interacts in a dense dielectric medium like glacial ice, the resulting particle shower generates a sub-nanosecond impulsive radio pulse through the Askaryan effect \citep{askaryan1962, zas1992, gorham2007}. Because polar ice is highly transparent to radio-frequency signals, boasting attenuation lengths on the order of a kilometer \citep{aguilar2022attenuation}---roughly ten times longer than for the optical Cherenkov light targeted by detectors like IceCube \citep{aartsen2013ice}---it is possible to construct sparse arrays approaching $100\,\rm km^3$ in volume. This massive scale is strictly required to reach the EeV energy thresholds---sitting well above the TeV to PeV astrophysical flux measured by IceCube---targeted by in-ice observatories like the Askaryan Radio Array (ARA) \citep{allison2012, allison2020} and the Radio Neutrino Observatory in Greenland (RNO-G) \citep{aguilar2021}, alongside balloon-borne NASA missions like the Antarctic Impulsive Transient Antenna (ANITA) \citep{gorham2018anita3} and the Payload for Ultrahigh Energy Observations (PUEO) \citep{abarr2025}. Launched in December 2025, PUEO completed an approximately 23-day flight over Antarctica, collecting over $50\,\rm TB$ of data comprising more than 100 million events. Despite their different operational strategies, all of these experiments share the same core task: isolating extremely rare, impulsive radio signals from an overwhelming background dominated by thermal noise \citep{dasgupta2023icrc}, anthropogenic contamination, and triboelectric effects \citep{aguilar2023tribo}.

Ultimately, while the scientific goals of technosignature searches and UHE neutrino astronomy are entirely different, both fields face the same needle-in-a-haystack problem \citep{wright2018}. This paper explores that shared methodological space. Using a synthetic wideband radio example, we demonstrate how technosignature searches can directly benefit from data-cleaning tools developed for UHE neutrino pipelines, provided they are carefully adapted to preserve potential SETI candidates.

\section{The Shared Rare-Event Detection Problem}
In SETI, candidate lists are often overwhelmed by terrestrial RFI; the BLC1 event is a prime example, where a compelling signal was only identified as local interference after an exhaustive, multi-stage vetting process \citep{sheikh2021}. UHE neutrino pipelines face a near-identical struggle against thermal noise and persistent narrowband contamination \citep{gorham2018anita3,allison2020}. Interestingly, the two fields operate with a perfect role reversal: the narrowband signals that SETI researchers search for are the very ``nuisance'' interference that neutrino analysts work to filter out. This irony motivates a unified analysis philosophy. Rather than forcing every event into a rigid signal-versus-background decision, pipelines should prioritize conservative, traceable cleaning and model-agnostic anomaly ranking. By preserving unusual residuals and maintaining rich metadata, both communities can ensure that a potential discovery—whether a cosmogenic neutrino or a technosignature---is not rejected simply because it failed to fit a pre-defined template.

\section{Proof-of-Concept: Conservative CW Mitigation for Rare-Event Searches}
\label{sec:cw_demo}

We illustrate the preserve-then-rank workflow using a fully synthetic
wideband radio example. Experiment-specific calibration, antenna response,
and trigger logic are not included; a full instrument-level validation is
deferred to future work.

Classical radio technosignature searches already have mature methods for rejecting radio-frequency interference (RFI), including ON--OFF source cadences and Doppler drift rate filtering \citep{enriquez2017}, as well as frequency occupancy and RFI environment characterization in later wideband analyses \citep{price2020,jacobsonbell2025}. The purpose of this paper is to address the following question: can a CW mitigation technique used in UHE neutrino analyses be adapted to a technosignature workflow if it is made conservative enough to avoid erasing uncataloged narrowband candidates?

In radio UHE neutrino experiments, narrowband continuous wave (CW) contamination is a dominant non-impulsive background. Persistent tones, primarily from station communications, satellite communications, and weather balloons (radiosondes), can mask the impulsive broadband radio signals expected from UHE neutrino-induced Askaryan emission. To preserve livetime, ARA analyses employ time-domain sine-subtraction techniques, originally developed by the ANITA Collaboration~\citep{gorham2018anita3} and later adopted and refined in ARA~\citep{allison2022a5}.

In this approach, candidate CW frequencies are first identified in the frequency domain by scanning the power spectrum for narrow peaks whose power exceeds a predefined threshold relative to the local spectral baseline. Both known persistent lines and newly appearing or drifting tones are flagged using power ratio tests and/or phase stability criteria. For each identified frequency $f_{\rm CW}$, a sinusoid of the form
$$S(t) = a \sin(2\pi f_{\rm CW} t) + b \cos(2\pi f_{\rm CW} t)$$
is then fit directly in the time domain, where $a$ and $b$ are free parameters that encode the amplitude and phase of the contaminating tone. The best-fit sinusoid is subtracted from the waveform. This procedure is repeated for all flagged frequencies in the event. By performing the subtraction in the time domain rather than applying a frequency-domain notch filter, the method minimizes ringing artifacts and preserves the shape of short broadband impulses.

For technosignature searches, however, this sine-subtraction method must be applied with care. A narrowband feature removed as CW contamination in a standard neutrino analysis is morphologically identical to the narrowband signal class historically targeted by radio SETI. In actual ARA and ANITA analyses, CW identification is often semi-blind: pipelines scan the spectrum and remove any narrowband features exceeding a pre-determined power threshold. A blind rule like this could easily eliminate a legitimate extraterrestrial signal of interest. Therefore, the strategy adopted here is strictly catalog-based: we subtract only frequencies that are independently known to be persistent anthropogenic contaminants, preserving uncataloged narrowband structure for later anomaly ranking.

\begin{figure}[!t]
\centering
\includegraphics[width=0.820\textwidth]{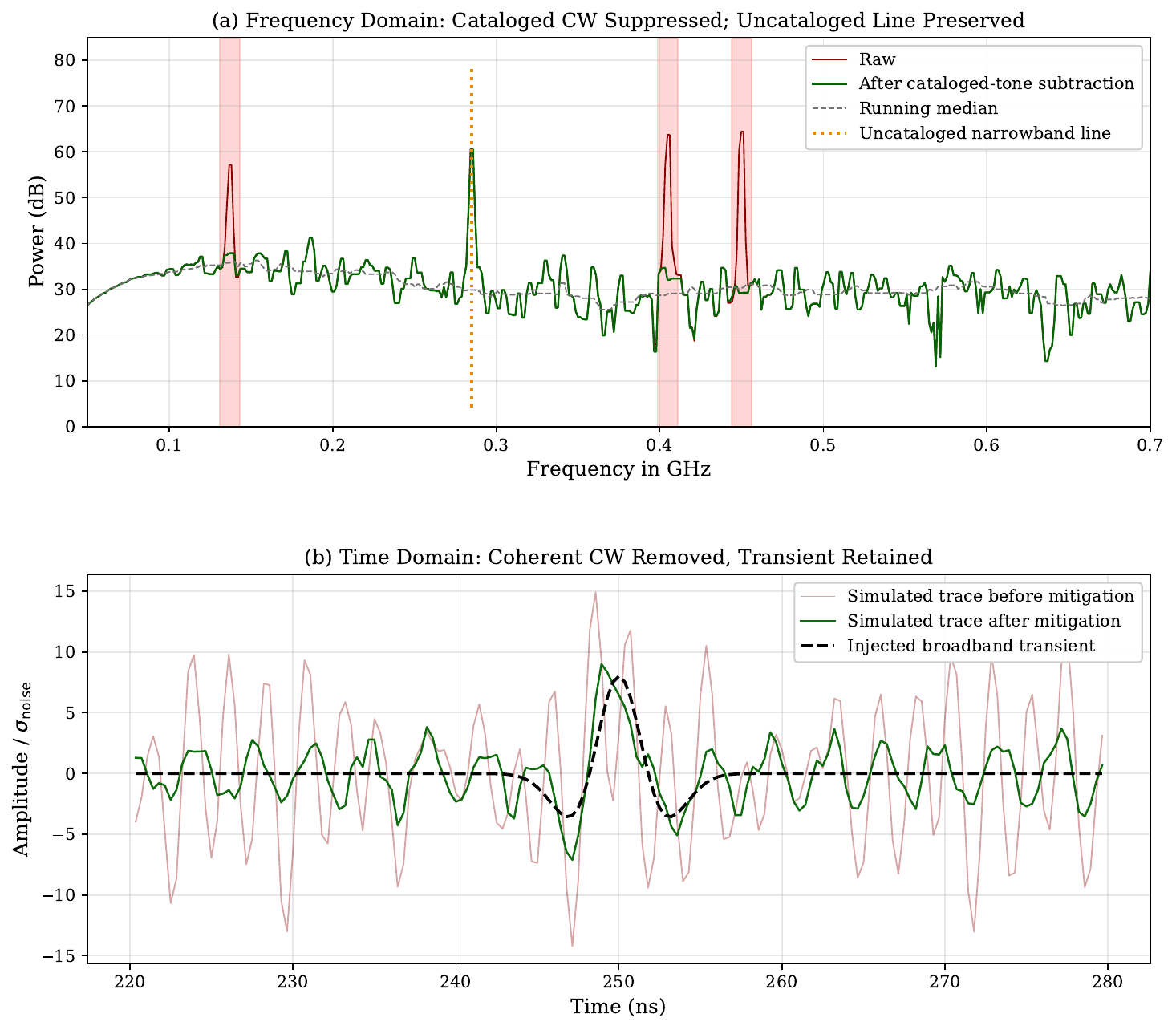}
\caption{Conservative CW mitigation demonstrated on a fully synthetic wideband radio dataset with realistic detector features. 
\textbf{(a)} Frequency spectrum before and after mitigation. 
Red shaded regions indicate the three cataloged anthropogenic CW tones at 137, 405, and 450\,MHz (representing satellite downlinks, weather balloon telemetry, and South Pole station communications, respectively). 
These tones are removed via least-squares time-domain sine subtraction as described in the text. 
The orange dotted line marks an uncataloged narrowband feature at 285\,MHz that is successfully preserved. 
\textbf{(b)} Time domain trace of the same waveform shown in panel (a), zoomed in to show the effect of mitigation around an injected nanosecond-scale broadband transient (dashed black line). The raw simulated trace before mitigation (light red line) is heavily obscured by CW interference. After subtraction of the cataloged tones, the mitigated trace (solid green line) shows significantly reduced coherent CW contamination, successfully recovering the underlying broadband impulsive signal and preserving both its original shape and temporal structure.}
\label{fig:cw}
\end{figure}

Figure~\ref{fig:cw} illustrates this idea using a fully synthetic wideband radio trace that incorporates realistic detector features. The synthetic thermal background is generated in the frequency domain by drawing Rayleigh-distributed spectral amplitudes with random phases, utilizing a smooth spectral envelope $\sigma(f)$. This envelope includes a soft band-limited response, a broad low-frequency excess, and a small gain ripple term, producing a non-flat but controlled wideband noise spectrum. The final voltage trace is normalized to a unit thermal noise RMS, meaning all injected components are expressed in units of $\sigma_{\rm noise}$ rather than calibrated detector voltage.

To test the catalog-driven mitigation, three tones are injected at 137\,MHz (satellite communications), 405\,MHz (weather balloon radiosonde telemetry), and 450\,MHz (South Pole station communications). These frequencies reflect persistent anthropogenic narrowband contamination commonly observed in ARA data \citep{allison2020,allison2022a5}. For each cataloged tone, the best-fit sinusoid is determined via least-squares and subtracted in the time domain, consistent with recent ARA analyses \citep{dasgupta2023icrc,dasgupta2024arena}.

An additional uncataloged narrowband tone is injected at 285\,MHz. This signal is deliberately excluded from the mitigation catalog so that it survives the cleaning process, serving as a proxy for an unknown narrowband candidate of potential technosignature interest. Finally, a nanosecond-scale Ricker pulse \citep{ricker1953} is injected at 250\,ns to serve as a proxy for a broadband impulsive transient.

Panel~(a) of Figure~\ref{fig:cw} shows the frequency-domain result of this procedure. The cataloged CW tones (red shaded regions) are strongly suppressed after sine subtraction, while the uncataloged narrowband feature at 285\,MHz (orange dotted line) is successfully preserved. Spectral power is shown in decibels (dB) with arbitrary normalization. In a technosignature search, such an uncataloged narrowband residual is not automatically classified as background; it could represent an unknown intermittent emitter, a drifting transmitter, or a candidate requiring further vetting.

Panel~(b) of Figure~\ref{fig:cw} shows the time-domain trace corresponding to the same event. Before mitigation, the trace is entirely dominated by coherent CW oscillations. After the time-domain subtraction of the known cataloged tones, the injected broadband transient is retained and clearly visible, demonstrating that the technique successfully filters interference without destroying impulsive physics targets.

The same preservation logic extends beyond the frequency domain. In the radio detection of UHE neutrinos, in-ice and balloon-borne experiments routinely use reconstructed arrival directions and event timing to remove anthropogenic backgrounds: events recurring from a fixed direction, or clustered tightly in time and space, are flagged as local emitters. Radio technosignature searches apply analogous spatial and temporal vetoes via ON--OFF cadences, direction-of-origin filters, and multibeam coincidence checks \citep{enriquez2017,pinchuk2022,zhang2020}. Recently, unsupervised methods such as GLOBULAR have further formalized this approach in the spectral domain by using HDBSCAN to cluster common RFI morphologies while preserving isolated outliers \citep{jacobsonbell2025}.

To the author's knowledge, the explicit formulation of a joint-feature workflow shared between UHE neutrino and technosignature analyses has not been presented before. Building on this symmetry, the core methodological contribution of this paper is to propose ranking residual events in a unified, multi-dimensional joint-feature space---incorporating time, frequency, bandwidth, duration, arrival direction, and polarization. In this framework, events forming high-density clusters within this feature space are cataloged as background RFI, while isolated outliers are retained as candidates for further investigation. This model-agnostic approach simultaneously suppresses the false-positive rate and minimizes the risk of false negatives, ensuring that morphologically novel technosignatures are not erroneously discarded.

\section{Background-Only Anomaly Ranking}
\label{sec:anomaly_ranking}

Conservative CW mitigation is the necessary first stage of the proposed workflow, but it leaves us with a critical question: once the obvious contaminants are removed, which residual events are actually worth following up? This is exactly where technosignature and UHE neutrino searches converge. Both fields face the same needle-in-a-haystack problem, operating under severe class imbalance where rare signals are hidden in overwhelmingly dominant noise. Analysis pipelines tuned aggressively to reduce false positives carry the risk of discarding genuine anomalies simply because they fail to match a predefined template.

To address this, we propose a second stage in the analysis pipeline based on background-only anomaly ranking. Instead of training a model to identify a specific potential signal, we let it learn the detailed characteristics of the background directly from the cleaned data. Each residual event is then assigned a score based on how poorly it fits that background profile. Because the model is tuned to the typical noise distribution, ordinary background events naturally receive low scores. On the other hand, events with anomalous spectral features, unexpected shapes, or rare waveform properties get a high anomaly score. This step effectively clears out the remaining background noise, leaving us with the actual anomalies. By sorting these outliers by their score, we get a straightforward shortlist of targets to look at next.

This classification is driven by metrics such as autoencoder reconstruction error or latent space clustering \citep{ma2023, jacobsonbell2025}. By automating the most tedious part of the search, we ensure that events the model simply cannot explain away as background float to the top. These outliers are then ready for the final stage of analysis: cross-correlation with RFI catalogs or detailed follow-up.

Importantly, we intend for this approach to complement our existing analysis frameworks, not replace them. In the UHE neutrino community, supervised machine learning techniques---most notably Linear Discriminant Analysis (LDA), rooted in Fisher's linear discriminant \citep{fisher1938}---remain the established benchmarks for identifying well-modeled impulsive Askaryan signals \citep{allison2022a5,dasgupta2023icrc}. These methods have been central to recent ARA analyses, where a 10\% subset of the full in-ice radio data is used to train an LDA classifier to maximize the efficiency of diffuse neutrino searches \citep{allison2022a5, dasgupta2024arena}.

Technosignature candidate events, however, are inherently unpredictable. We could be looking for a drifting narrowband carrier, a complex wideband burst, or even something as exotic as a directed neutrino beam from a distant civilization---analogous to those produced by particle accelerators at CERN or Fermilab. In a landscape this wide open, a model-agnostic anomaly score acts as a vital safety net. It keeps our search flexible, ensuring we do not accidentally discard a true discovery simply because it failed to match our preconceived signal expectations.

Translating this ``preserve-then-rank'' philosophy into practice is entirely feasible using the modern techniques discussed above \citep{jacobsonbell2025}. Ultimately, the technical bridge between our communities comes down to metadata preservation. By keeping rejected events accessible so they can be tested under new assumptions later, we essentially future-proof our pipeline against our own analytical blind spots.

\section{Discussion and Conclusions}
\label{sec:conclusions}

This work is based on the shared analytical techniques between technosignature searches and ultra-high energy neutrino astronomy. Despite having different scientific goals, the two communities have independently arrived at remarkably similar analysis strategies: they both carefully remove persistent interference, preserve potential candidates at every step, and rank the remaining events based on how anomalous they appear. These parallels stem from a shared fundamental challenge---detecting rare signals buried in complex, heavily contaminated backgrounds. We make this connection explicit through three main contributions.

First, we showed that catalog-based time-domain sine subtraction---a technique implemented by the ANITA Collaboration and later adopted in ARA analyses \citep{gorham2018anita3,allison2022a5}---can be effectively adapted for wideband technosignature searches when limited to documented persistent contaminants. Applied conservatively, this step simultaneously improves the visibility of broadband impulsive transients for particle astrophysics and preserves uncataloged narrowband features for technosignature vetting. As illustrated in Figure~\ref{fig:cw}, a single processing decision can serve both science goals from the same data stream.

Second, we identified a structural equivalence between the spatiotemporal consistency tests used in UHE neutrino experiments and RFI rejection methods in radio SETI. Although developed separately, both approaches follow the same density-based logic: clustered events are classified as background, while isolated events are retained as candidates. By combining these into a unified joint feature space---including direction, arrival time, frequency, bandwidth, duration, and polarization---we provide a natural, instrument-agnostic framework for cross-community anomaly ranking.

Third, we proposed that background-only anomaly ranking is the most suitable second stage in the preserve-then-rank workflow, precisely because technosignature morphologies are unconstrained. While UHE neutrino analyses can use supervised classifiers optimized for the Askaryan signal \citep{allison2022a5,dasgupta2025}, technosignature searches lack an equivalent template. A model-agnostic anomaly score therefore acts as an important safety net, sensitive to any deviation from the learned noise distribution, including exotic possibilities such as directionally structured artificial neutrino emission \citep{learned2008,learned2009}.

The results presented here are based on synthetic data and serve as a proof-of-concept rather than a sensitivity study. Future work includes applying anomaly-ranking algorithms to shared synthetic benchmarks, testing the joint feature framework on public radio technosignature datasets, and eventually validating the ``preserve-then-rank'' pipeline on real wideband radio data from both communities.





\section*{Acknowledgments}
I thank the organizing committee of IAUS\,404: \textit{Advancing the Search for Technosignatures} for the invitation to present this work. The talks, poster sessions, and vibrant discussions during the symposium further shaped the ideas and methods presented here. My interest in the cross-disciplinary synergies between UHE-neutrino experiments and technosignature searches was first sparked at the ``Bridging Multi-Messenger Astronomy and SETI: The Deep Ends of the Haystack'' workshop at The Ohio State University. I am deeply grateful to Brian Lacki for his continued guidance and for connecting me with the broader technosignature community, to Vishal Gajjar for valuable discussions, and to John F.\ Beacom for his encouragement and insightful comments that substantially improved this work. I also thank Amy Connolly for her helpful feedback and encouragement. Finally, I acknowledge my colleagues in the ARA, ANITA, and PUEO collaborations for the foundational pipeline expertise and collaborative experience that underlie the perspective presented here. This work was generously supported by the Center for Cosmology and AstroParticle Physics (CCAPP) at The Ohio State University.

\end{document}